\documentclass{article}

\usepackage{arxiv}

\usepackage[utf8]{inputenc} 
\usepackage[T1]{fontenc}    
\usepackage{url}            
\usepackage{booktabs}       
\usepackage{amsfonts}       
\usepackage{nicefrac}       
\usepackage{microtype}      
\usepackage{lipsum}		
\usepackage{xr}       
\usepackage{hyperref}
\usepackage{graphicx}
\usepackage{natbib}
\usepackage{doi}
\usepackage{gensymb}
\usepackage{siunitx}

\usepackage{amsmath}
\usepackage{comment}

\title{Hyperspectral Imaging for Inline Quality Monitoring of Roll-to-Roll Slot-Die Coated Organic Photovoltaic Active Layers}

\author{ \href{https://orcid.org/0000-0003-3820-3324}{\includegraphics[scale=0.06]{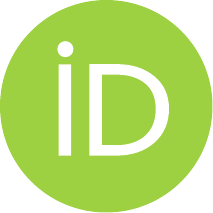}\hspace{1mm} Abdelouadoud Mammeri} \\
	Department of Energy, Technical University of Denmark\\
	Denmark\\
	\texttt{abdmam@dtu.dk} \\
    	\And
	\href{https://orcid.org/0000-0001-9245-8198}{\includegraphics[scale=0.06]{orcid.pdf}\hspace{1mm}Søren Alkærsig Jensen} \\
	Danish Fundamental Metrology A/S\\
	Denmark\\
    \texttt{saj@dfm.dk} \\
    	\And
	\href{https://orcid.org/0009-0007-8404-3460}{\includegraphics[scale=0.06]{orcid.pdf}\hspace{1mm}Søren AR Kynde} \\
	Danish Fundamental Metrology A/S\\
	Denmark\\
    \texttt{srk@dfm.dk} \\
    	\And
	\href{https://orcid.org/0000-0003-4719-2356}{\includegraphics[scale=0.06]{orcid.pdf}\hspace{1mm}Astrid Tranum Rømer} \\
	Danish Fundamental Metrology A/S\\
	Denmark\\
    \texttt{asr@dfm.dk} \\
    \And
	\href{https://www.iscent.fi/}{\includegraphics[scale=0.06]{orcid.pdf}\hspace{1mm}Raimo Korhonen} \\
	M-Boss OY\\
	Finland\\
    \texttt{raimo.korhonen@m-boss.fi}
    \And
	\href{https://orcid.org/0009-0008-7708-1845}{\includegraphics[scale=0.06]{orcid.pdf}\hspace{1mm}Matteo Ciambezi} \\
    Independent Researcher\\
	Sweden\\
    \texttt{matteo.ciambezi@gmail.com} \\
    \And
	\href{https://orcid.org/0000-0003-4843-7474}{\includegraphics[scale=0.06]{orcid.pdf}\hspace{1mm}Moises Espindola} \\
    F.Junckers Industrier\\
    Denmark\\
    \texttt{mespindola01@hotmail.com} \\
	\And
	\href{https://orcid.org/0000-0002-3145-0229}{\includegraphics[scale=0.06]{orcid.pdf}\hspace{1mm}Jens Wenzel Andreasen} \\
	Department of Energy, Technical University of Denmark\\
	Denmark\\
    \texttt{jewa@dtu.dk} \\
}

\renewcommand{\shorttitle}{\textit{arXiv} Template}

\hypersetup{
pdftitle={A template for the arxiv style},
pdfsubject={q-bio.NC, q-bio.QM},
pdfauthor={David S.~Hippocampus, Elias D.~Striatum},
pdfkeywords={First keyword, Second keyword, More},
}

\begin{document}
\maketitle

\begin{abstract}
Lab-scale organic solar cells have shown high efficiency and strong potential for large-scale production. Among the fabrication techniques for upscaling production, roll-to-roll slot-die coating offers a cost-effective route toward scalable manufacturing. However, reliable in-line metrology is needed for real-time assessment and quality control of the coating during roll-to-roll manufacturing. In this work, we employ a hyperspectral camera for inline monitoring of P3HT and OIDTBR active layers in a roll-to-roll setup. Using multivariate curve resolution, we reliably resolved the spatial distribution of donor and acceptor contributions in the coating. Furthermore, by fitting the absorbance spectra of P3HT, we map the exciton bandwidth, providing insights into the polymer's aggregation and chain interactions. These results demonstrate the potential of hyperspectral imaging as a powerful tool for real-time monitoring and quality control in large-scale organic solar cell production.
\end{abstract}

\keywords{Solar cells \and Real time monitoring  \and Conjugated polymers \and Non fullerene acceptor \and Spatial mapping}

\section{Introduction}

Harnessing renewable energy sources is an important focus of the current century. Solar energy, an inexhaustible and accessible energy source, is a great candidate. Organic Solar Cells (OSCs) offer several advantages, including low cost, low weight, simple preparation methods, flexibility, and suitability for large-area fabrication. The laboratory scale
devices (\SI{1}{\centi\metre\squared} photoactive area) are around \SI{20}{\percent} Power Conversion Efficiency (PCE), and large area devices are around 15\% \cite{basu2024large}. However, a key challenge in large-scale OSC fabrication lies in an effective large-scale production method. To overcome this challenge, precise management of all parameters involved in OSC fabrication is necessary, including ink composition, coating method (e.g., spin-coating, blade-coating, slot-die coating), drying kinetics, and film thickness. In recent years, several inline and in situ characterization techniques have been adapted and developed for this purpose.\\ \\
These techniques span different directions, covering a range of methods to characterize the quality and performance of OSCs. A Grazing Incident Small Angle X-ray Scattering (GISAXS) setup was used to unveil the structure and ordering of inline coated films with a speed of around \SI{0.5}{\metre\per\minute} \cite{sorensen2021situ,10.1063/1.4892526}.
Laser reflectometry was utilized to study the drying process of blade-coated OSC active layers \cite{SCHMIDTHANSBERG2011509}.
Ellipsometry was also employed to measure the thickness and the dielectric constant of the different layers in OSC at \SI{10}{\metre\per\minute} web speed and \SI{2}{\milli\metre} spot width \cite{LOGOTHETIDIS2013144}. Light Beam Induced Current (LBIC) with a scanning reflected laser was employed to obtain a 2D current map of OSC at a \SI{1.2}{\metre\per\minute} web speed and a spatial resolution of  $\SI{0.25}{\milli\metre}\times\SI{1}{\milli\metre}$ \cite{https://doi.org/10.1002/adom.201470028}. A combination of ellipsometry, Raman and photoluminescence techniques were used for an inline, real-time monitoring of P3HT:PCBM based solar cells \cite{zetaalphachialpharhoiotaacutealphadeltaetavarsigma2016situ}\\
Among optical metrology techniques, transmission UV--Vis spectroscopy is widely used to monitor the evolution of absorption spectra during OSC film formation, providing insight into molecular aggregation, polymer ordering, and intra-chain coupling. Several studies have employed \textit{in situ} transmission measurements to investigate drying, aggregation, and morphological evolution during spin coating, blade coating, and slot-die coating of organic semiconductors and bulk heterojunction blends~\cite{C3TC32077D,https://doi.org/10.1002/smtd.202100585,D3TC02562D,https://doi.org/10.1002/adma.202105114,https://doi.org/10.1002/aenm.202303661,doi:10.1021/acs.jpcc.2c06337,https://doi.org/10.1002/solr.202000086,doi:10.1021/acsami.0c12390}. However, these approaches rely on single-point measurements, with spot sizes ranging from tens of micrometers to few millimeters at most, providing insight into film heterogeneity.

Hyperspectral Imaging (HSI) is an advanced imaging technique that captures a full spectrum for every pixel in an image. By combining spatial information with the chemical specificity provided by the spectral data, HSI becomes a highly powerful analytical tool. It has been applied across a wide range of fields, including guided surgery \cite{lu2014medical}, detecting oral neoplasia \cite{10.1158/1940-6207.CAPR-11-0555}, agriculture \cite{RAM2024109037}, pollution and particulate monitoring \cite{s19143071}. Brianna Conrad \textit{et al.} \cite{10.1063/5.0131691} employed hyperspectral photo- and electroluminescence imaging (Photon etc.) to identify recombination defects in dilute-alloy InGaAs solar cells. They used a laser source of \SI{532}{\nano\metre} with an image size of approximately 18 $\times$ \SI{18}{\milli\metre^2}. The wavelength range was \qtyrange{1000}{1600}{\nano\metre}, and the spectral resolution was around \qty{2}{\nano\metre}. Similarly, Gilbert El-Hajje \textit{et al.} \cite{C6EE00462H} utilized hyperspectral photo- and electroluminescence imaging (Photon etc.) to spatially map the charge carrier transport efficiency in perovskite solar cells, achieving a spatial resolution of \qty{2}{\micro\metre}. In another study, Matthew P. Peloso \textit{et al.} \cite{10.1063/1.3664134} also used hyperspectral electroluminescence imaging (P\&P Optica PPO-HYPSPEC-001) in the wavelength range of \qtyrange{950}{1250}{\nano\metre} to characterize the minority carrier diffusion length in multicrystalline silicon solar cells.

In this work, we demonstrate the use of a hyperspectral camera in transmission mode with a telecentric optics to monitor and quantify the quality of donor and acceptor active layers in organic solar cells fabricated by slot-die coating for inline roll-to-roll manufacturing. This approach enables spatially resolved, real-time analysis of the moving films by multivariate curve resolution and fitting the absorbance spectra, thereby providing detailed insight into film quality in terms of thickness uniformity, chemical species homogeneity, polymer aggregation and even the drying process.  
\section{Experimental}
\subsection{Setup}

The setup used for this paper was a hyperspectral measurement system in transmission mode. The light source was a Tungsten-Halogen Light Source (SLS201L Thorlab) coupled to an optical fiber. The light guided by the fiber was projected onto a parabolic mirror to collimate it before it reached the sample. The hyperspectral camera was a Pika XC2 (Resonon) equipped with a telecentric lens (0.377X, 4/3" C-Mount TitanTL Edmund optics). Both were integrated onto a roll-to-roll machine to perform the inline measurements. 

The hyperspectral camera is a line-scan camera that measures one-dimensional spatial images, with the second dimension of the detector used to capture decomposed spectral data as a function of wavelength. It captures a line image \textit{frame} of the moving web, which is perpendicular to the film's direction of motion; this spatial dimension is denoted as \textit{Track direction}. Combining consecutive frames yields a 2-dimensional spatial image with spectral (transmission) information at every pixel. The exposure time $\tau$ was fixed at \SI{12}{\milli\second}. The spatial direction along the web's movement was denoted as the \textit{Moving direction}. The final measurements consisted of a continuous frame capturing the moving web, from which we obtained a hyperspectral cube. The measurements generate 2.82 MiB per frame. 
The transmission of the coated films $\left(T(\lambda)\right)$ was obtained by the following equation:
\begin{equation}
    T(\lambda)=\frac{I_{\textrm{meas}}(\lambda )- I_{\textrm{dark}}(\lambda )}{I_{\textrm{ref}}(\lambda )- I_{\textrm{dark}}(\lambda )}
\end{equation}
where the measured frame intensity is $I_{\textrm{meas}}$, the substrate measured intensity is $I_{\textrm{ref}}$ and the background light is $I_{\textrm{dark}}$.

\subsection{Materials}

The materials used in this study include Poly(3-hexylthiophene-2,5-diyl) (P3HT) with a regioregularity of \SI{97.6}{\percent} and a molecular weight of \SI{60.15}{k\dalton}, sourced from Ossila, and (5Z,5'Z)-5,5'-((7,7'-(4,4,9,9-tetraoctyl-4,9-dihydro-s-indaceno[1,2-b:5,6-b']dithiophene-2,7-diyl)bis(benzo[c][1,2,5]thiadiazole-7,4-diyl))bis(methanylylidene))bis(3-ethyl-2-thioxothiazolidin-4-one) (OIDTBR) from One Material. Chlorobenzene with 99.8\% purity (Sigma-Aldrich) was used as a solvent to dissolve P3HT and OIDTBR, yielding a final solution concentration of \SI{10}{\milli\gram\per\milli\litre} for each dye. All chemicals were used as received, without further purification or treatment. 

\section{Results and discussion}
\subsection{Chemical composition ratio}
We coated on PET a mixture of P3HT and OIDTBR at different pump rates using PEEK Mixing Tees (Low Pressure Cheminert by VICI JOUR). The two solutions of \SI{10}{\milli\gram\per\milli\litre} concentration were pumped independently using two identical syringe pumps. The P3HT solution was initially pumped at \SI{20}{\micro\litre\per\minute} for \SI{5}{\minute}. Its flow rate was then decreased in steps of \SI{5}{\micro\litre\per\minute} every \SI{2}{\minute} until it reached \SI{0}{\micro\litre\per\minute}. Simultaneously, after the first \SI{5}{\minute}, pumping of the OIDTBR solution was initiated at \SI{5}{\micro\litre\per\minute} and increased in \SI{5}{\micro\litre\per\minute} increments every \SI{2}{\minute} until reaching \SI{20}{\micro\litre\per\minute}. After a total pumping time of \SI{11}{\minute}, the P3HT flow was stopped, while OIDTBR continued to be pumped at \SI{20}{\micro\litre\per\minute} for an additional \SI{5}{\minute}. The coating speed was \SI{30}{\cm} per \SI{}{\minute}. We employed Multivariate Curve Resolution (MCR) \cite{DEJUAN202159,camp2019pymcr} to quantify the concentration of P3HT and OIDTBR. MCR analysis required initialization using reference spectra for both P3HT and OIDTBR. To ensure consistency with the experimental measurements, the initial spectra of P3HT and OIDTBR were acquired using the same camera system employed throughout the study. These reference spectra are provided in the Supplementary Information. The MCR takes these spectra and linearly decomposes each frame of $400$ (spatial) $\times$ $924$ (spectral) pixels to produce $400 \times 2$ and $2 \times 924$ matricies. The first matrix represents the track spatial concentration (contribution) of the two elements. The second matrix represents the reconstructed spectra. The MCR method produces the concentration as a scaling factor of the P3HT and OIDTBR spectra. To obtain meaningful numbers, we normalize them. 

\begin{figure}
    \centering
    \includegraphics[width=1\linewidth]{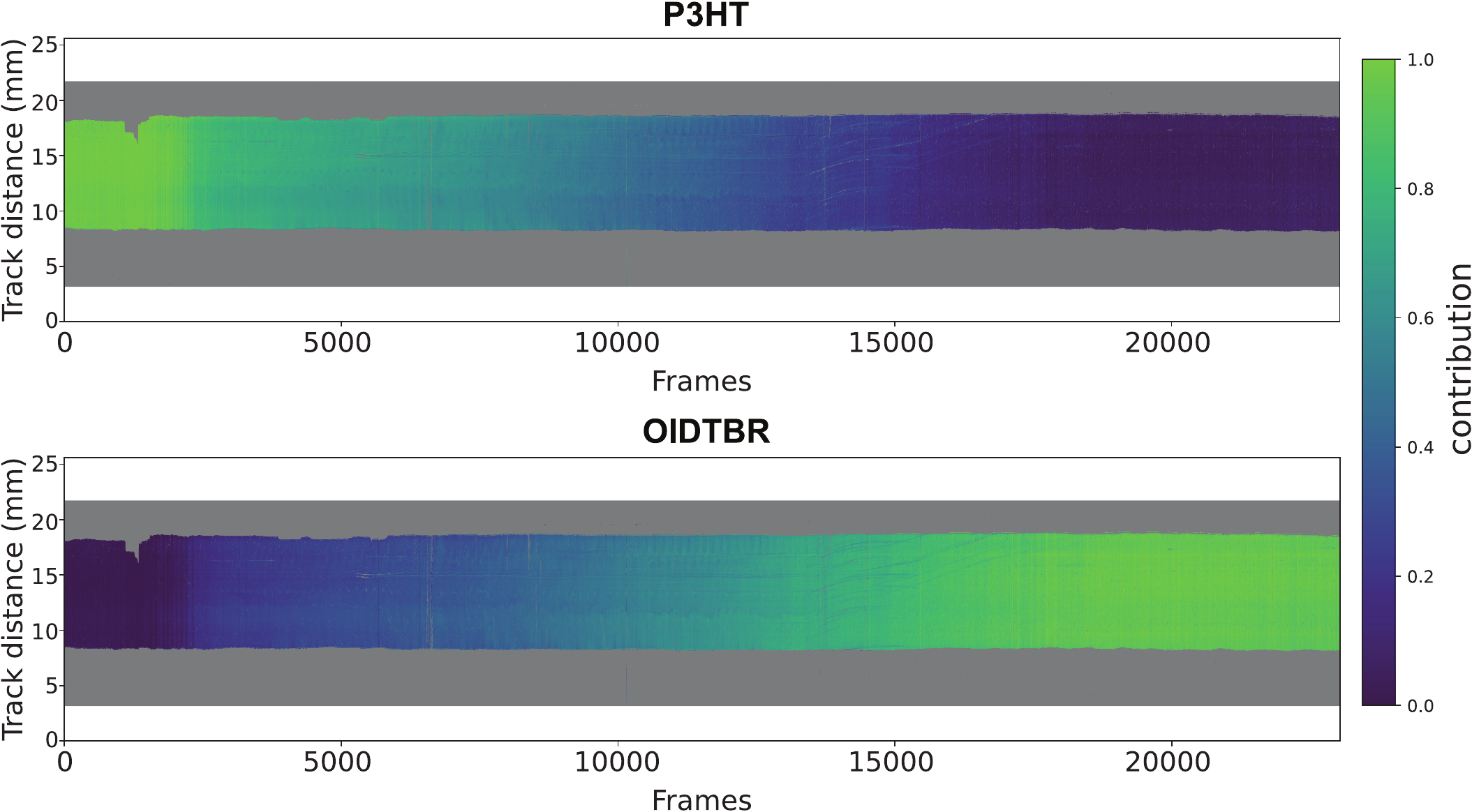}
    \caption{Normalized P3HT and OIDTBR contribution. Gray pixels represent data retained after applying a 97.5\% goodness-of-fit threshold.}
    \label{P3HT_MCR_framses}
\end{figure}
\begin{figure}[h]
    \centering
    \includegraphics[width=1\linewidth]{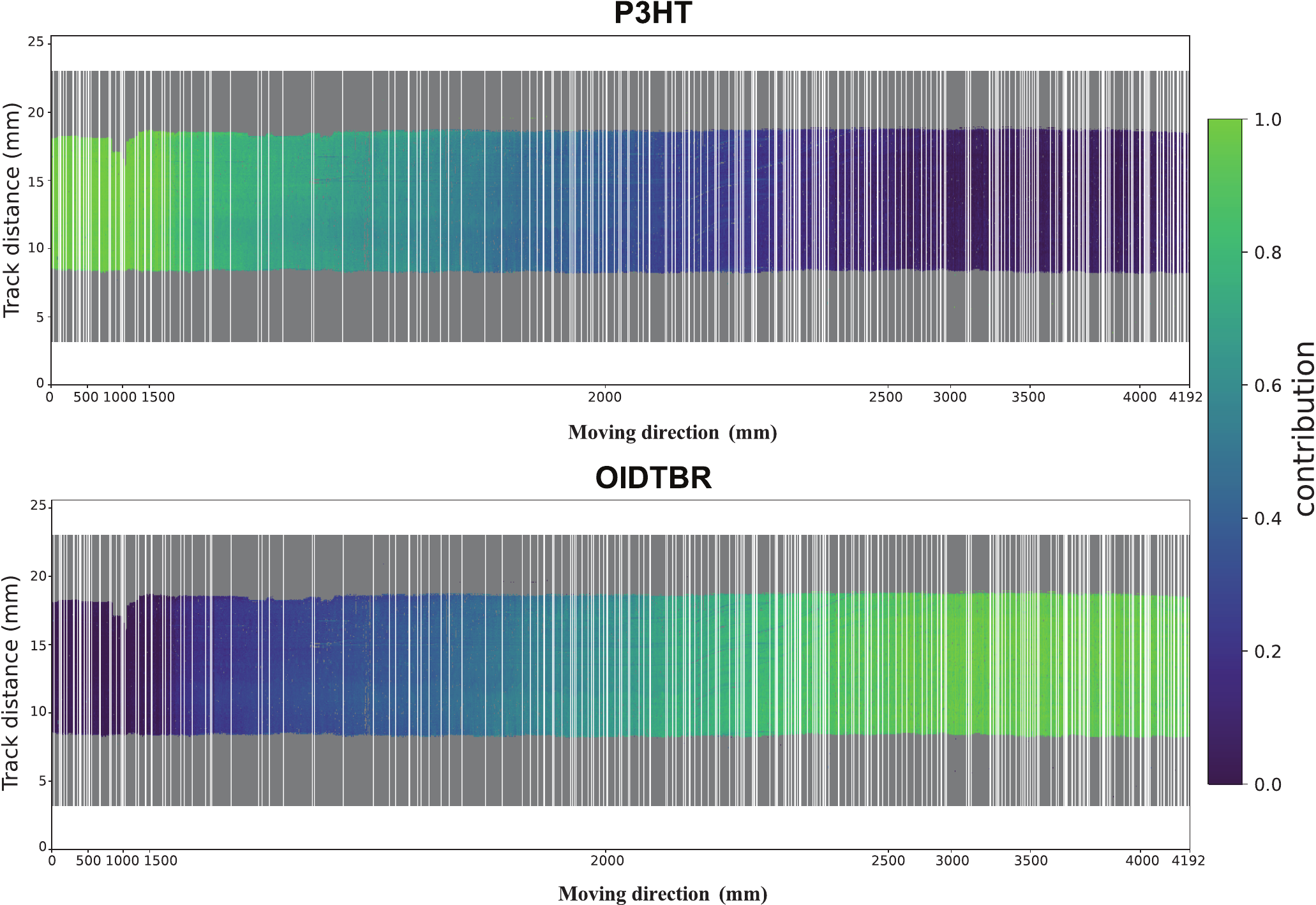}
    \caption{P3HT and OIDTBR normalized contribution plotted in physical positions. The white spacing is uncollected data of the moving web while captured frames are being processed. The white-gap width $g$ is compressed using the transformation $g_{plot}=s\log(1+g/s)$, where $s=0.005$ is the compression scale and $g_{plot}$ is the plotted gap.}
    \label{MCR_R2R_real_distance}
\end{figure}

\raggedbottom
Figure \ref{P3HT_MCR_framses} shows the accumulated frames of the normalized P3HT concentration. The analysis was performed in real time using parallel computing with three processing agents. The camera integration time was only \SI{12}{\milli\second}, allowing frames to be acquired much faster than they could be processed. To accommodate this mismatch, up to 600 frames were buffered in RAM during processing. Once the buffer reached its maximum capacity, newly acquired frames were discarded until processing caught up with the acquisition rate, resulting in frame loss. To represent the true spatial position of the measurements, the number of discarded frames was taken into account when calculating distance, and the resulting P3HT and OIDTBR contributions are plotted in Figure \ref{MCR_R2R_real_distance}.

As shown in Figure S1, the MCR algorithm converged rapidly when both components were present, requiring, on average, approximately \SI{200}{\milli\second} per frame. In contrast, processing times exceeded \SI{20}{\second} in regions containing only P3HT or only OIDTBR. This slower convergence arose because the algorithm attempted to reconstruct the absent component. Although the fitted concentration of the absent component remained low, the reconstructed spectrum in these cases was a noisy and distorted version of the initial spectrum rather than a meaningful spectral contribution. An example of the reconstructed and measured spectra, along with the initial P3HT and OIDTBR spectra, is presented in S2. The spatial and spectral goodness-of-fit metrics are presented in Figures S3 and S4, respectively.

Figure S5 shows the Spectrally Averaged Normalized Root Mean Square Error. There, we can see that most of the reconstruction discrepancies are coming from the PET substrate. In the same manner, Figure S6 shows the Spatially Averaged Normalized Root Mean Square Error. Initially, there is a low spectral related error between \SI{400} and \SI{650}{nm} because the P3HT spectrum exhibits well-defined features in this range. After introducing OIDTBR, we also present its features up to around \SI{750}{nm}, thereby further reducing the spectral error range. 
\subsection{Structural ordering and excitonic bandwidth analysis of P3HT}

\begin{figure}[H]
    \centering
    \includegraphics[width=1\linewidth]{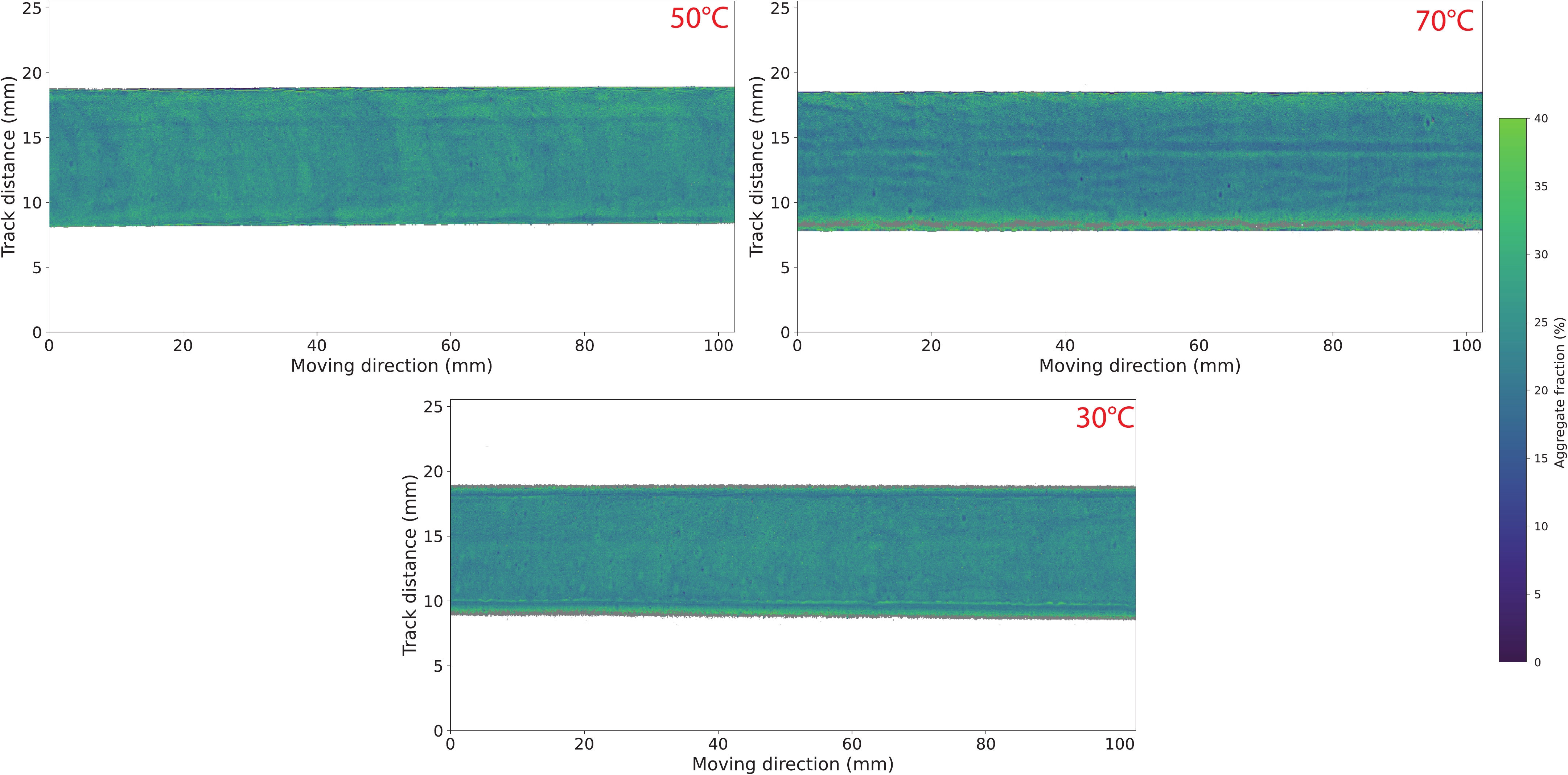}
    \caption{The crystallinity fraction of P3HT in the films according to: $\frac{\int aggregation}{ \int aggregation + 1.39 \times \int amorphous}$. The 1.39 factor accounts for the fact that amorphous P3HT absorbs less than the aggregated P3HT \cite{10.1063/1.3110904}. The Gray pixels have a goodness-of-fit below 93\%. White pixels were filtered out based on absorbance, since they represent areas without film coverage.}
    \label{aggregation_fraction}
\end{figure}

To quantify the the aggregation of P3HT coated films, the Spano's model was used \cite{10.1063/1.1914768,chu2024low}: 
\begin{equation}
\label{eq:Absfit}
A(E)=\sum_{m=0}^{4}e^{-S}\frac{S^m}{m!}\left(1-\frac{W\,e^{-S}}{2E_p}\sum_{n=1}^{4}\frac{S^{n}}{n!\,(n-m)}\right)G\!\left(E - E_{0-0}- mE_p,\Gamma\right) +G\left(E-E_{amorph},\Gamma_{amorph}\right)
\end{equation}

where \(S=1\) is the Huang--Rhys factor, \(W\) is the excitonic bandwidth, and \(E_P=\SI{0.179}{\eV}\) corresponds to the C=C stretching mode that determines the energy separation between the 0--0 and 0--1 transitions \cite{10.1063/1.3110904}. \(G\) is a Gaussian lineshape function with broadening \(\Gamma\) centered at  $E_{0-0}+ mE_p$, while \(\Gamma_{\mathrm{amorph}}\) represents the Gaussian broadening associated with the amorphous P3HT contribution. Finally, \(E_{0-0}\) is the energy of the 0--0 transition. During the fitting procedure, \(m\) was fixed at 4, meaning that only the first four vibronic transitions were considered. The effect of the excitonic bandwidth on the vibronic transitions of P3HT is illustrated in Figure S7, which shows spectra from two pixels with different excitonic bandwidths. The coated films were moved at \SI{50}{\centi\metre\per\minute}, which translates to a \SI{100}{\micro\metre} spatial resolution in the moving direction. In the track direction, the resolution is \SI{64}{\micro\metre}, meaning that the transmission signal is the average over such an area. The P3HT film was coated using a solution flow rate of \SI{30}{\micro\litre\per\minute} and a coating speed of \SI{30}{\centi\metre\per\minute}.
\begin{figure}[H]
    \centering
    \includegraphics[width=1\linewidth]{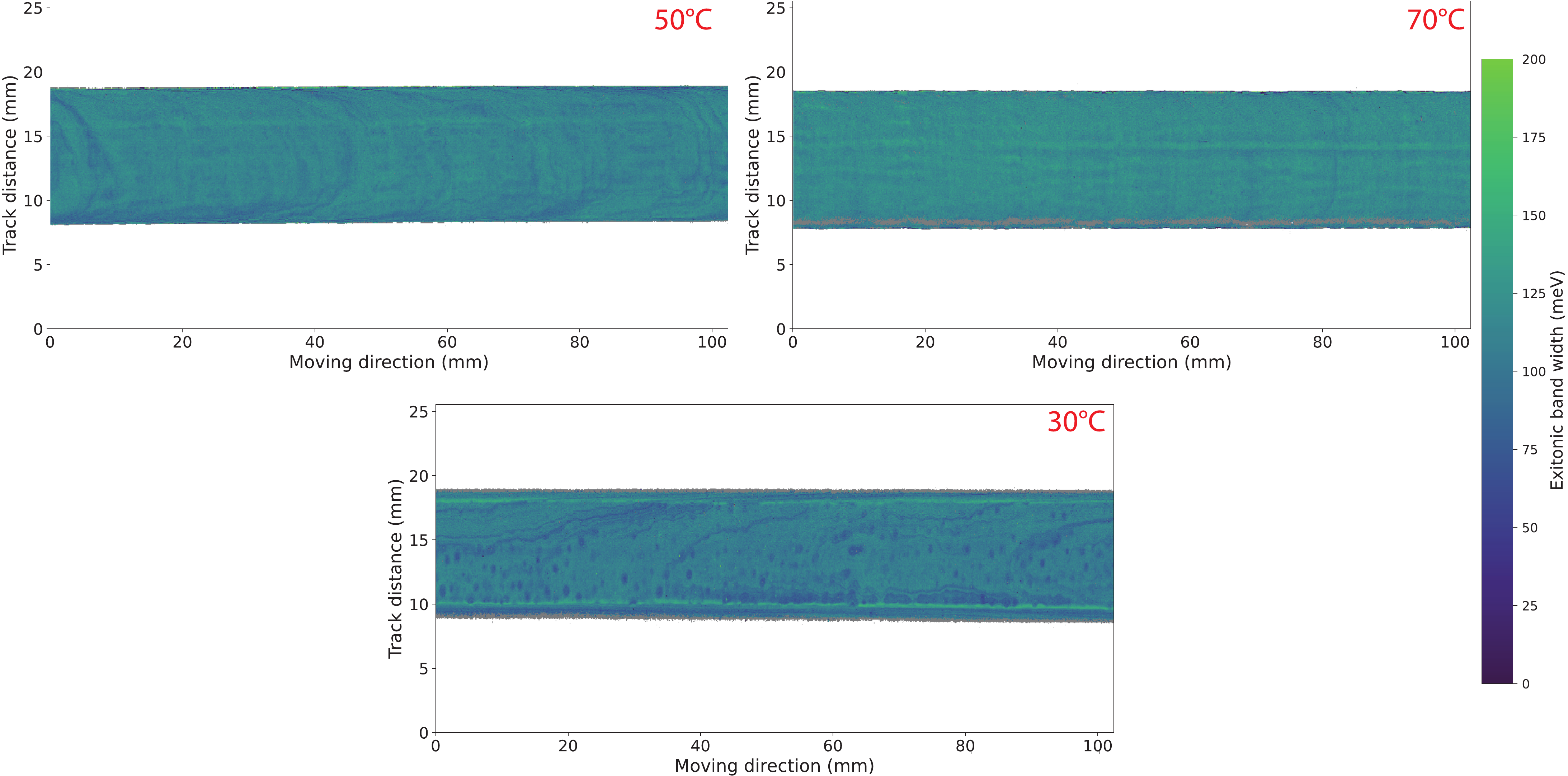}
    \caption{The excitonic band width spatial map for the coated P3HT at 30, 50 and \SI{70}{\celsius}. The Gray pixels have a goodness-of-fit below 93\%. The white pixels were excluded based on their absorbance values, as they correspond to regions where no film was present.}
    \label{w}
\end{figure}
Spatially resolved maps of the excitonic bandwidth $W$ (Figure \ref{w}) and aggregation fraction (Figure \ref{aggregation_fraction}), extracted from hyperspectral imaging through Spano-model fitting, reveal a strong influence of coating temperature on the uniformity of molecular ordering within the slot-die coated P3HT films. The temperature-dependent behavior of the coated films reflects the interplay between chlorobenzene evaporation and polymer chain mobility, which are the primary factors governing film formation in this case. As solvent evaporation drives the transition from solution to solid, polymer chains simultaneously diffuse and self-organize until solidification, with the competition between these processes determining the final morphology and homogeneity \cite{cummings2018modeling,panzer2024unified}. If solvent evaporation occurs too slowly, concentration gradients, convective flows and drying-front instabilities develop within the wet film, resulting in spatially heterogeneous film formation as in the case of the \SI{30}{\celsius} coated film. Conversely, if solvent evaporation is sufficiently rapid while chain mobility remains high enough to permit local structural relaxation, a more uniform microstructure can be established before the system vitrifies  \cite{10.1039/c6ee01623e,https://doi.org/10.1002/solr.202300349}. The parabola-like pattern in the excitonic bandwidth map and aggregation fraction is due to the pinning and depinning of the wet-dry film interface \cite{shiratori2025critical,wang2015molecular}. These maps provide a reliable tool for quantitative spatial quality control of polymers such as P3HT. We also tested real-time fitting of Spano's model (not shown in this work), which required, on average, \SI{30}{\milli\second} per pixel, corresponding to approximately \SI{3.2}{\second} per frame (with a spatial binning factor of 4). This processing time could be significantly reduced with parallel computation using a GPU. 

Figure S8 represents the thickness deviation across the film. The Pearson correlation between the latter and the aggregation fraction values was -0.402, -0.269, and -0.423 for the 30, 50, and \SI{70}{\celsius} coated films, respectively, which is reasonable as the drying process underlays both quantities.
The relative error of the excitonic band, shown in Figure S9, ranges from 6\% to 14\%, with most values clustered around 10\%. Although the uncertainty is somewhat elevated, the results remain sufficiently accurate to provide a quantitative assessment of the energetic disorder in P3HT.
The absolute error of the aggregation fraction is presented in Figure S10. The error ranges from 1.5 to 5 \% across the films.

\section{Conclusion}
In summary, this paper demonstrates the use of a hyperspectral camera in transmission mode as a quantitative tool for assessing the quality of the roll-to-roll slot-die active layer. Using multivariate curve resolution, the spatial contributions of P3HT and OIDTBR gradient coating can be reliably identified with minimal spatial and spectral errors. The system is also capable of identifying the aggregation ratio and the excitonic bandwidth of P3HT, providing insights into the polymer's energetic and structural disorder. 

\section*{Competing Interests}
The authors declare that they have no known competing financial interests or personal relationships that could have appeared to influence the work reported in this paper.
\section*{Acknowledgments}
The authors would like to thank FOM Technologies for supporting this work. 
\section*{Data availability}
Data presented in this paper and the supplementary materials may be obtained from the authors upon request.
\section*{Supplementary information}  
The supplementary materials are included with the paper. 

\section*{Ethical approval}
This article does not contain any studies with human or animal subjects.
\section*{Funding}
This work was partly funded by Innovation Fund Denmark (IFD) under File no. 3109-00024B, Eureka Eurostars project E2897 QualSurf. 

\bibliographystyle{unsrt}
\bibliography{references}  

\end{document}


\subsection*{Chemical composition ratio}
\begin{figure}[H]
    \centering
    \includegraphics[width=0.7\linewidth]{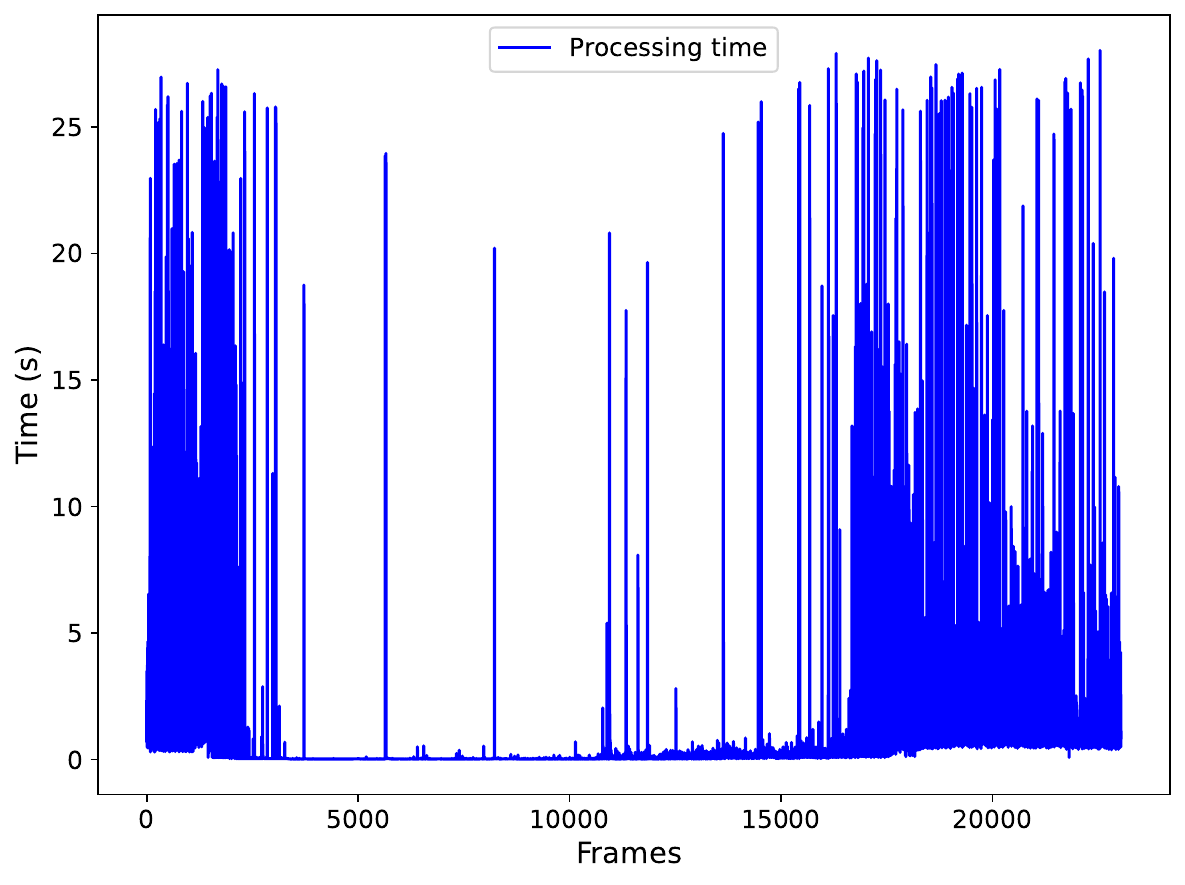}
    \caption{Processing time of each frame during the MCR measurement.}
    \label{MCR_processing_time}
\end{figure}
\begin{figure}[H]
    \centering
    \includegraphics[width=0.7\linewidth]{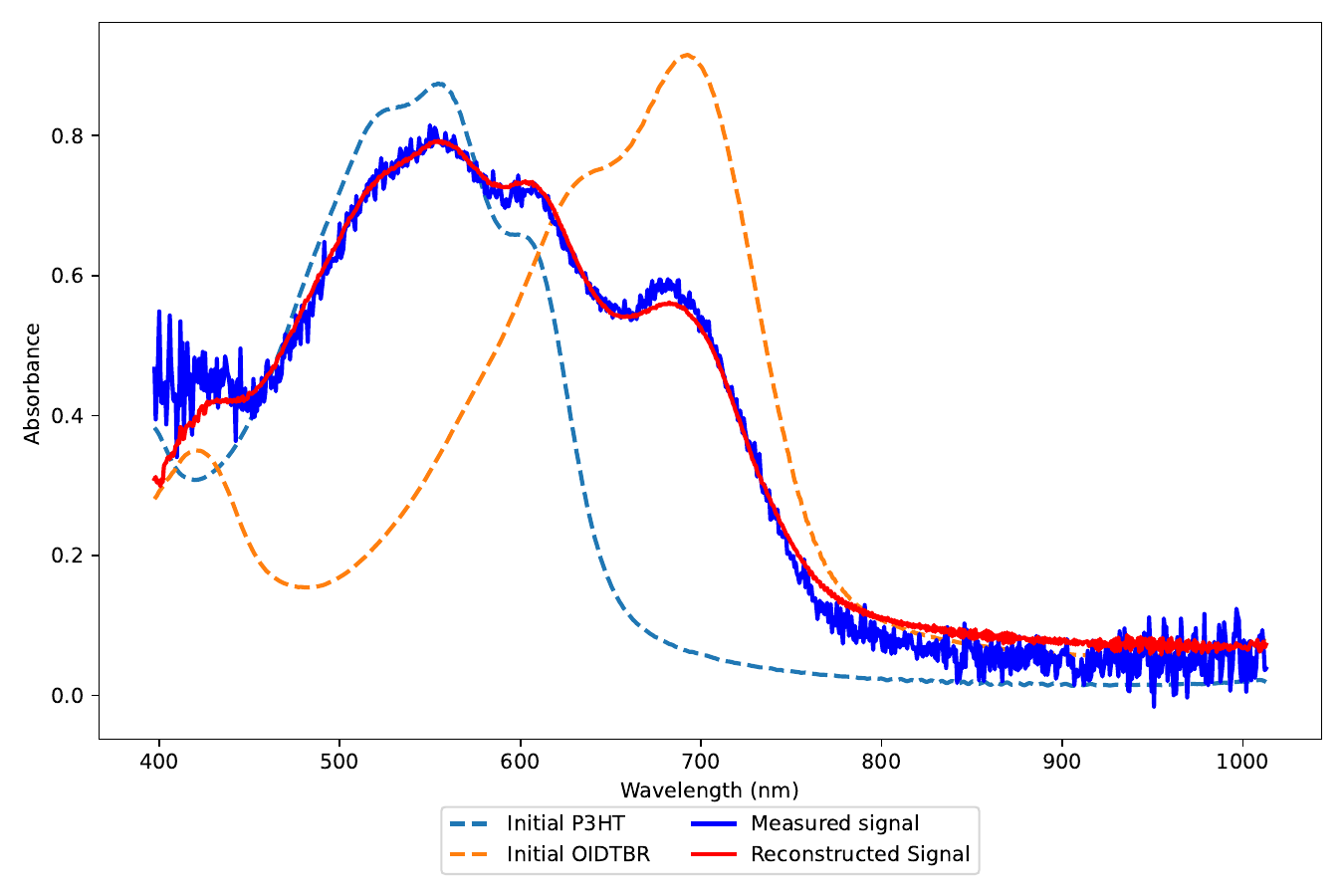}
    \caption{Initial spectra of P3HT and OIDTBR alongside an example of the measured and reconstructed signal.}
    \label{MCR_reconstructed}
\end{figure}
\begin{figure}[H]
    \centering
    \includegraphics[width=1\linewidth]{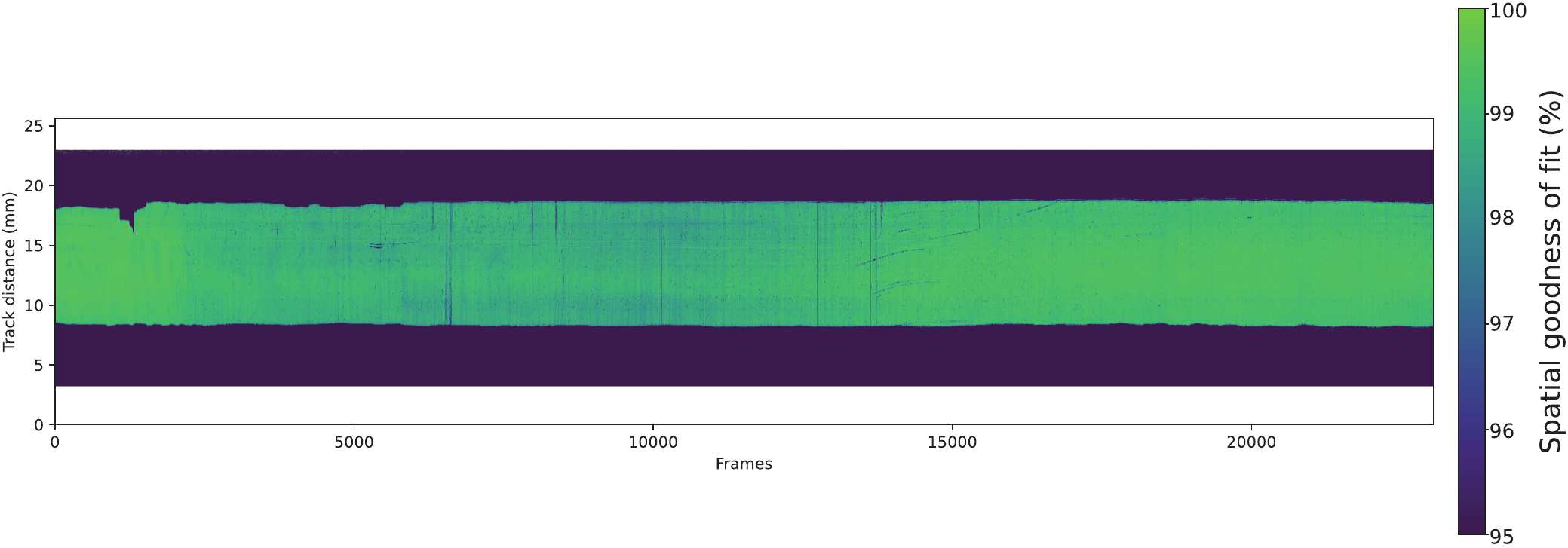}
    \caption{Spatial goodness of fit: : 1-$\frac{{\sqrt{\langle Residual^2\rangle}}_{spectral}}{{\sqrt{\langle measurement^2\rangle}_{spectral}}}$.}
    \label{MCR_GOF_spatial}
\end{figure}

\begin{figure}[H]
    \centering
    \includegraphics[width=0.9\linewidth]{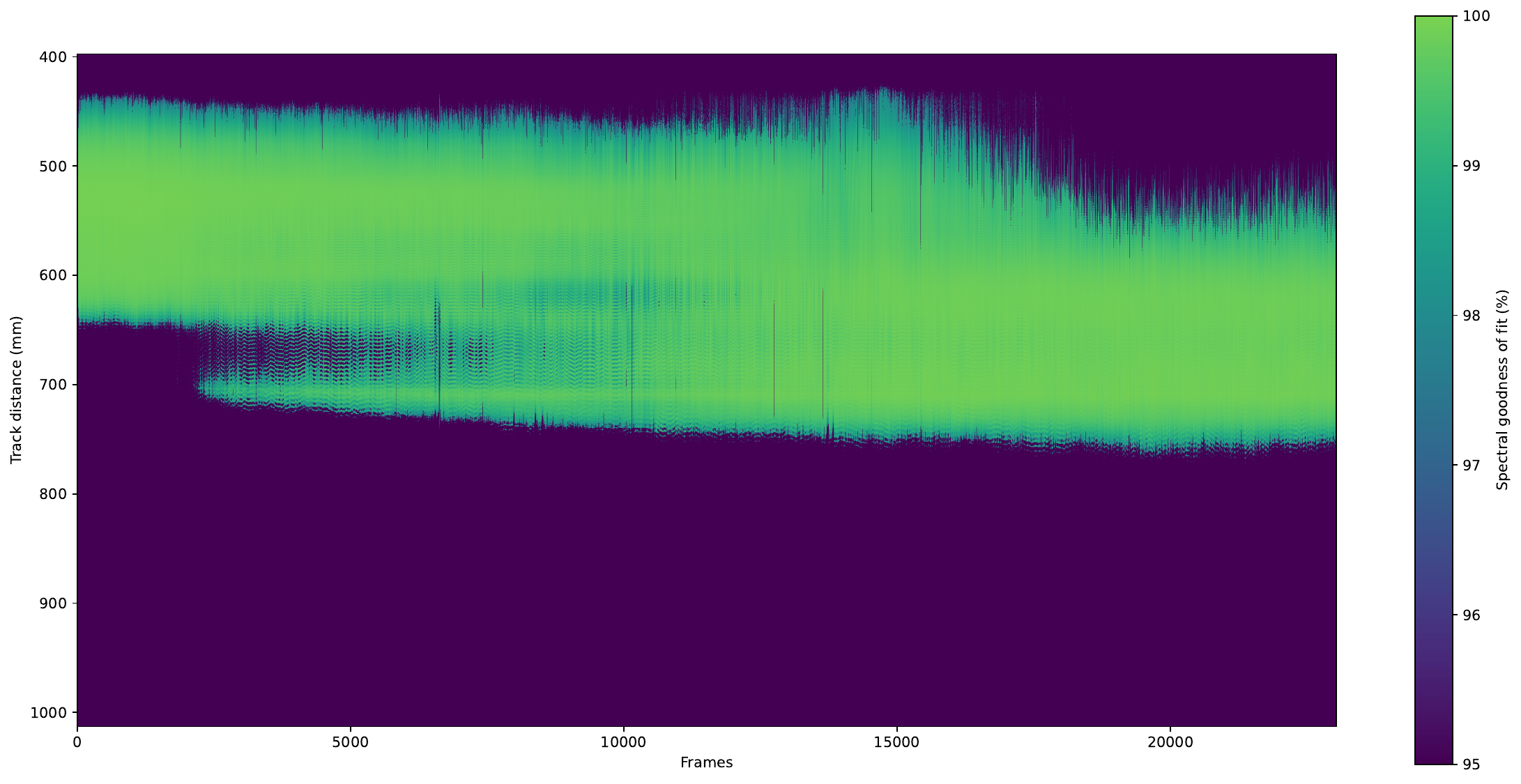}
    \caption{Spectral goodness of fit: 1-$\frac{{\sqrt{\langle Residual^2\rangle}}_{spatial}}{{\sqrt{\langle measurement^2\rangle}_{spatial}}}$.}
    \label{MCR_GOF_spectral}
\end{figure}
\begin{figure}[h]
    \centering
    \includegraphics[width=0.9\linewidth]{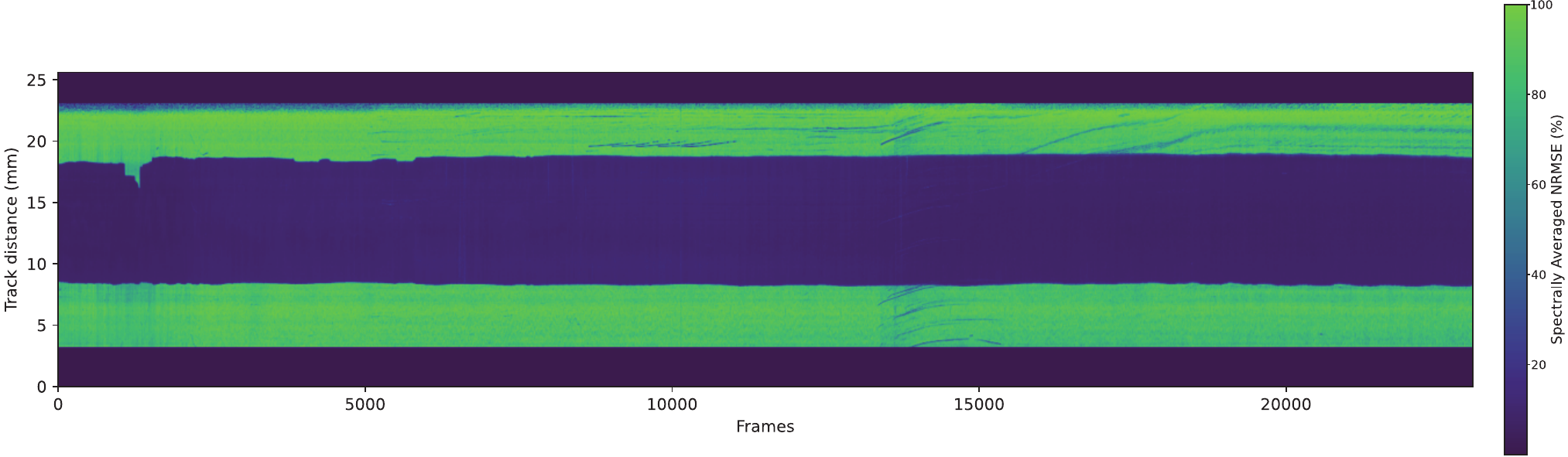}
    \caption{Specterly Averaged Normalized Root Mean Square Error: $\frac{{\sqrt{\langle Residual^2\rangle}}_{spectral}}{{\sqrt{\langle measurement^2\rangle}_{spectral}}}\times 100$.}
    \label{MCR_spatial_residual}
\end{figure}

\begin{figure}[H]
    \centering
    \includegraphics[width=1\linewidth]{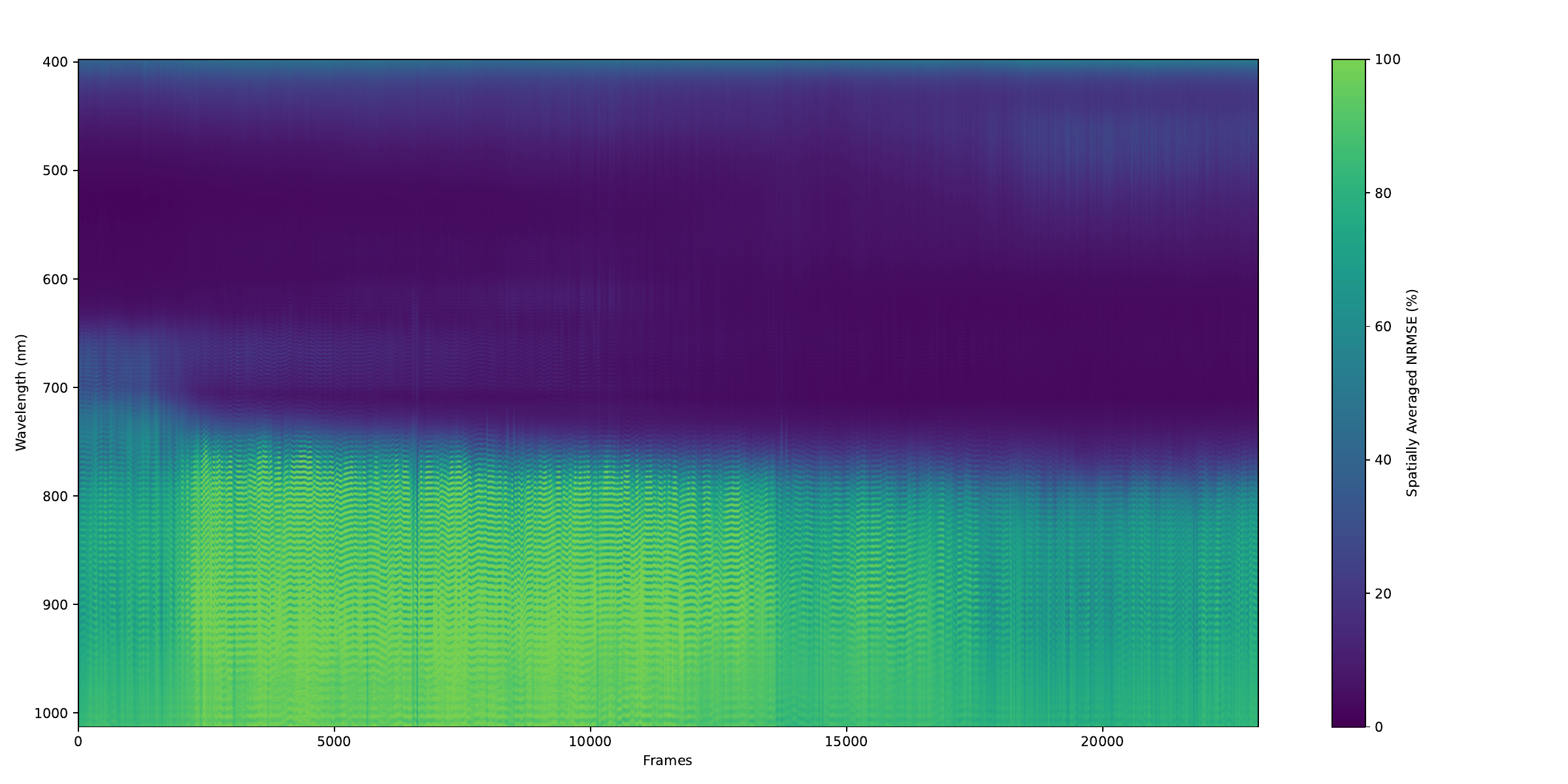}
    \caption{Spatially Averaged Normalized Root Mean Square Error: $\frac{{\sqrt{\langle Residual^2\rangle}}_{spatial}}{{\sqrt{\langle measurement^2\rangle}_{spatial}}}\times 100$.}
    \label{MCR_spectral_residual}
\end{figure}

\subsection*{Structural ordering and excitonic bandwidth analysis of P3HT}

\begin{figure}[H]
    \centering
    \includegraphics[width=1\linewidth]{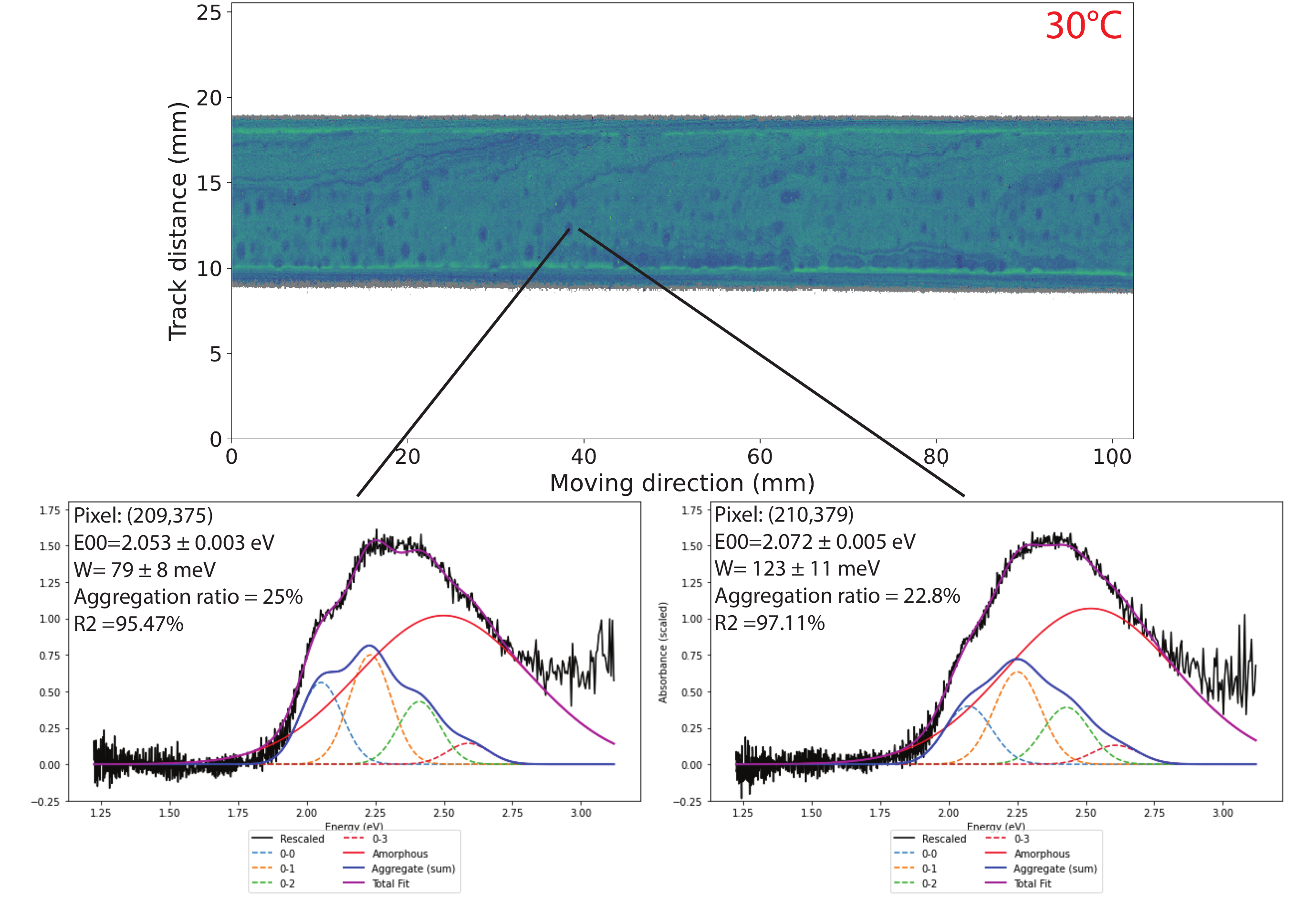}
    \caption{Two example spectra with low and high excitonic bandwidth are shown for pixels $(209,375)$ and $(210,379)$, respectively. The spectrum at pixel $(209,375)$ exhibits a stronger $0$--$0$ transition relative to the $0$--$1$ transition, resulting in a larger excitonic bandwidth ($W$). This behavior is associated with more ordered P3HT aggregates and longer exciton coherence lengths. In contrast, the spectrum at pixel $(210,379)$ shows a weaker $0$--$0$ transition and a more pronounced $0$--$1$ transition, indicating a lower degree of molecular order.}
    \label{2_fit_examples}
\end{figure}

\begin{figure}[H]
    \centering
    \includegraphics[width=1\linewidth]{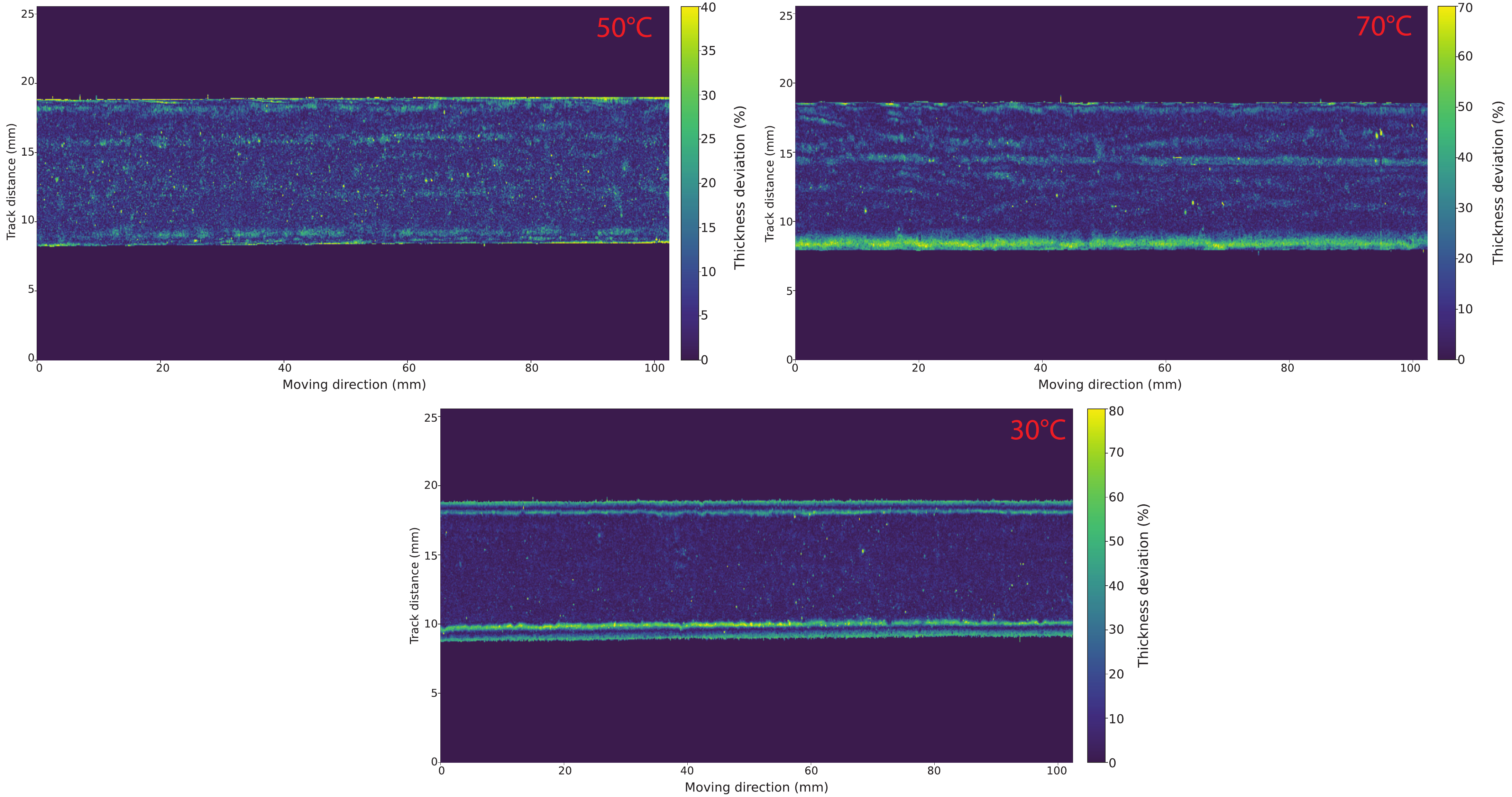}
    \caption{Thickness homogeneity of the coated P3HT at different temperatures calculated with the following equation: $d= \frac{\lvert\langle log(T(\lambda)) \rangle_{\lambda}-\langle log(T(\lambda)) \rangle_{f,\lambda}\rvert}{\langle log(T(\lambda)) \rangle_{f,\lambda}}$, where $\langle \cdot \rangle_{\lambda}$ is the average over the spectral dimension of the captured frame, $f$ is all pixels belonging to the film region in the frame.}
    \label{STD_map}
\end{figure}

\begin{figure}[H]
    \centering
    \includegraphics[width=1\linewidth]{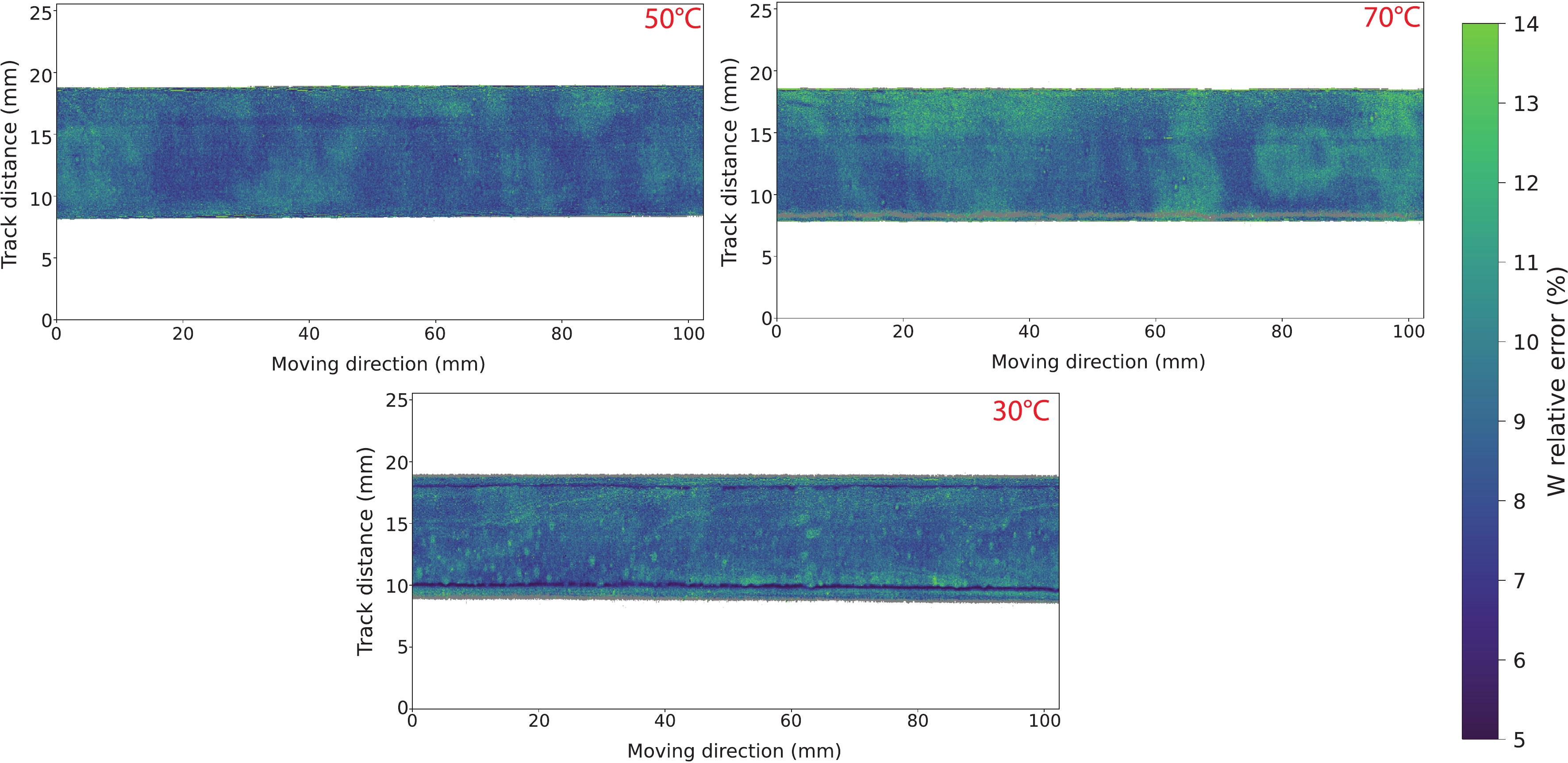}
    \caption{Excitonic band width relative error quantified through the covariance matrix.}
    \label{w_error}
\end{figure}

\begin{figure}[H]
\centering
\includegraphics[width=1\linewidth]{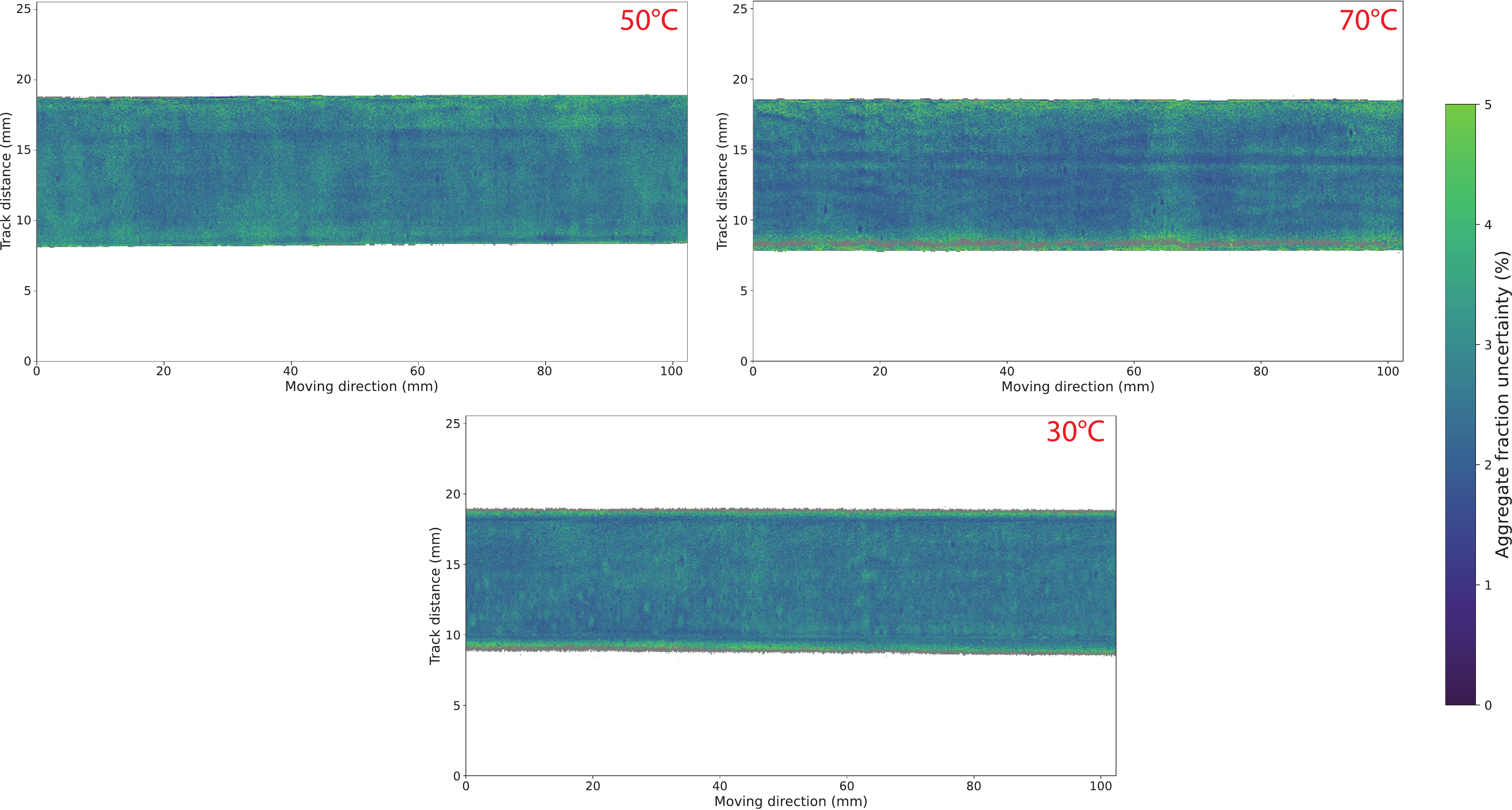}
\caption{The uncertainty in the aggregation fraction was evaluated using Monte Carlo uncertainty propagation. A total of 3000 parameter sets were sampled from a multivariate normal distribution with mean given by the fitted parameters and covariance matrix. For each sampled parameter set, the aggregate and amorphous spectra were reconstructed, and the aggregation fraction was recalculated from the corresponding peak areas. The standard deviation of the resulting aggregation fractions was taken as the uncertainty associated with the fitting procedure. This uncertainty was subsequently combined with the uncertainty of the correction factor ($\pm 0.10$ from reference 27 in the manuscript) using standard first-order (partial derivative) error propagation.}
\label{aggregation_error}
\end{figure}